\documentclass[lettersize,journal]{IEEEtran}
\usepackage{amsmath,amsfonts,amssymb}
\usepackage{stfloats}
\usepackage{url}
\usepackage{graphicx}
\usepackage{cite}
\usepackage{bbm}
\usepackage{physics}
\usepackage{mleftright}

\DeclareMathOperator*{\argmin}{argmin}
\DeclareMathOperator*{\argmax}{argmax}
\DeclareMathOperator{\diag}{diag}

\begin{document}

\title{A Simple Capacity-Achieving Scheme for Channels with Polarization-Dependent Loss}

\author{Mohannad~Shehadeh,~\IEEEmembership{Graduate Student Member,~IEEE}
	and~Frank~R.~Kschischang,~\IEEEmembership{Fellow,~IEEE}
	\thanks{Submitted on Aug.~8, 2022; revised version submitted on Nov.~3, 2022. \newline The authors are with the Edward S. Rogers Sr. Department of Electrical and Computer Engineering, University of Toronto, Toronto, ON M5S 3G4, Canada (emails: \{mshehadeh, frank\}@ece.utoronto.ca).
	}}

\maketitle

\begin{abstract}
We demonstrate, for a widely used model of channels with
polarization dependent loss (PDL), that channel capacity  
is achieved by a simple interference cancellation scheme in conjunction with a universal precoder. 
Crucially, the proposed scheme is not only 
information-theoretically optimal, but it is also 
exceptionally simple and concrete. It transforms the PDL
channel into separate scalar additive white Gaussian noise channels, allowing off-the-shelf coding and modulation schemes
designed for such channels 
to approach capacity. The signal-to-noise ratio (SNR) penalty incurred 
under 6 dB of PDL is reduced to the information-theoretic minimum of a mere 1 dB
as opposed to the 4 dB SNR penalty incurred under naive over-provisioning.
\end{abstract}


\section{Overview}

\IEEEPARstart{P}{olarization}-dependent loss (PDL) is a 
capacity-reducing impairment in 
polarization-division-multiplexed (PDM) 
coherent optical transmission systems \cite{shtaif,shtaif-ultimate,foschini,dumenil-dissertation,dumenil-fundamental}. 
PDL mitigation schemes have been studied in a flurry of recent works 
\cite{elaine,elaine2,dumenil-dissertation,dumenil-jlt,dumenil-fundamental,fujitsu,huawei,oyama,oyama-zf,damen,pairwise,china,awwad-jlt,awwad-dissertation} as well as older works \cite{shtaif-st,shtaif-st2,old-pt,scrambling,stokes-space}.
Renewed interest in PDL stems from its expected impact on next-generation
optical networks which are dense in PDL-inducing components such as
reconfigurable add-drop multiplexers (ROADMs) \cite{dumenil-dissertation,probabilistic-design}.
In this paper, we provide a careful information-theoretic analysis of PDL-impaired PDM
channels, modelling them as \textit{compound channels} \cite{compound-lapidoth} and, in accordance with these models,
provide a provably optimal, simple, low-complexity PDL mitigation scheme.

Throughout this paper, 
we assume a common memoryless 
model for a PDL-impaired PDM channel with no insertion loss (IL) uncertainty 
and with the channel
parameters being perfectly known to the receiver but unknown
to the transmitter. This model is considered in \cite{dumenil-dissertation,dumenil-jlt,
	huawei,damen,awwad-dissertation,old-pt,winzer-shtaif,oyama}.
Our goal is to find a scheme which maximizes the rate of reliable communication that can be \textit{guaranteed}, given a 
known \textit{worst-case} PDL value.
We provide a simple scheme that achieves this goal \textit{and} reduces the problem
to separate communication across scalar additive white Gaussian noise (AWGN) channels, so that standard coding and modulation 
schemes for such channels can be directly applied without loss 
of optimality.

Our results can be summarized as follows.
Given a class of channels with a \textit{worst-case} PDL value of
\begin{equation*}
	10\log_{10}\mleft(\frac{1+\alpha}{1-\alpha}\mright)\text{ dB}
\end{equation*}
where $\alpha \in [0,1)$ is fixed, a fundamental asymptotic SNR penalty
relative to a classical AWGN channel 
is incurred. This penalty depends on the extent to which the two polarizations
are jointly processed. In particular, the SNR penalty is, under
\begin{itemize}
	\item no joint coding or decoding,
	\begin{equation}\label{nj-penalty}
		10\log_{10}\mleft(\frac{1}{1-\alpha}\mright) \text{ dB;}
	\end{equation}
	\item joint coding with parallel and independent decoding,
		\begin{equation}\label{p-penalty}
			10\log_{10}\mleft(\frac{1}{1-\alpha^2}\mright) \text{ dB;}
		\end{equation}
	\item and joint coding with joint or successive decoding,
	\begin{equation}\label{sic-penalty}
		10\log_{10}\mleft(\frac{1}{\sqrt{1-\alpha^2}}\mright)  \text{ dB.}
	\end{equation}
\end{itemize}

The improved but suboptimal SNR penalty \eqref{p-penalty}
can be achieved by schemes along the lines of those described in \cite{pairwise,oyama}.
The optimal (smallest possible) SNR penalty \eqref{sic-penalty} is achieved by our proposed 
scheme which constitutes a simple precoder in combination 
with a linear minimum mean square error
(LMMSE) plus successive interference cancellation (SIC) receiver.
The precoder is judiciously chosen so that the effective \textit{channel matrix}
after interference cancellation is an \textit{orthogonal design} \cite{orthogonal-designs-book,orthogonal-designs-paper} in the channel parameters, 
i.e., is unconditionally orthogonal.
The resulting precoders in the cases of real- and complex-valued channel matrices
are essentially permutations of those considered in \cite{pairwise,oyama,huawei}, but it is precisely this correct choice
of permutation, or equivalently, interference cancellation order, 
that is vital to achieving \eqref{sic-penalty}.

The remainder of this paper is organized as follows.
In Section \ref{Existing Schemes}, we comment on relationships between our proposed scheme and analysis and existing work. 
In Section \ref{PDL-Cap}, we analyze the capacity of
a PDL-impaired channel as a \textit{compound channel} \cite{compound-lapidoth}
and establish some requisite background. 
We 
then provide capacity-achieving schemes in 
Sections \ref{Real-Scheme} and \ref{Complex-Scheme} for 
the cases of real- and complex-valued channel models, respectively, both of which are common in the PDL literature. 
In Section \ref{Practical}, we provide a coarse-grained performance analysis 
of the proposed scheme, some comments on practical considerations,
and suggestions for future work. We end with concluding remarks in Section \ref{Conclusion}.

\section{Existing Schemes}\label{Existing Schemes}

\subsection{LMMSE-SIC Schemes}

It is well-known in information theory that LMMSE-SIC schemes are 
capacity-achieving in a variety of multiple-input multiple-output (MIMO) AWGN channel
settings, provided that perfect interference cancellation is performed (refer to, e.g., \cite[Chapter~8]{tse-wireless-communication}). 
This is practically accomplished
by separately coding the transmitted data streams and performing the interference cancellation
after forward error correction (FEC). Such schemes, apart from avoiding infeasibly complex
joint maximum likelihood (ML) detection of data streams, effectively synthesize separate scalar sub-channels on which codes designed for scalar channels can be employed without loss of optimality. Such coded LMMSE-SIC schemes have also been
experimentally validated in space-division-multiplexed
(SDM) optical transmission systems \cite{USIC_1,USIC_2,USIC_3}
and thus are known to be practical.

Achievable information rates under 
LMMSE-SIC schemes were recently investigated by Chou and Kahn in \cite{elaine} 
in the general context of mode-dependent loss (MDL) in SDM systems including PDL-impaired PDM systems as a special case. 
In \cite{elaine}, it is noted that, as is the case with slow fading wireless channels,
such schemes are suboptimal under separate coding of data streams since the achievable rate
is limited by the capacity of the worst sub-channel whose identity is not known at the transmitter.

In this paper, we demonstrate how this limitation of LMMSE-SIC
can be bypassed in the special case of memoryless PDL
with no IL uncertainty. 
In particular, we demonstrate that there exists a 
universal precoder which symmetrizes the channel 
so that the sum of the worst-case capacities
of the sub-channels induced by LMMSE-SIC is equal to the worst-case sum of these
capacities, rendering such a scheme optimal.

\subsection{Precoding with ML Detection}

In \cite{awwad-dissertation,awwad-jlt,shtaif-st,shtaif-st2,old-pt,damen}, 
space--time coding schemes from wireless communication are adapted to produce
polarization--time codes which are typically paired with ML receivers. Moreover, in \cite{dumenil-dissertation,dumenil-jlt}, the authors consider 
precoding schemes which operate only across polarizations and in-phase and quadrature components to reduce the complexity of ML processing. 
Our work demonstrates
that under our modelling assumptions, such schemes are of no information-theoretic
benefit compared to simpler precoders along the lines of \cite{oyama,pairwise}
when combined with a carefully designed linear interference cancellation architecture
and codes designed for scalar AWGN channels. In particular, any apparent performance
differences are essentially shaping and coding gains that could
be relegated, without loss of optimality, to the outer AWGN channel codes. 

If, however, we include uncertainty in the polarization-average loss or IL in the channel model, as in \cite{foschini}, 
such schemes could, in principle, have better outage performance
than the proposed scheme. 
Stated more concretely, in the presence of random PDL \textit{and} random IL, 
adding an IL margin to the PDL margin \eqref{sic-penalty} 
so as to guarantee a certain outage probability does
not result in the smallest theoretically possible penalty---or 
equivalently, highest achievable information rate---for that
outage probability.
Indeed, there could exist schemes which 
enable reliable communication 
across capacity-equivalent channel realizations
having combinations of high IL with low PDL and low PDL
with high IL. We leave the quantification of what
gains are left on the table in such a setting as
a question for future work.

\subsection{Information-Theoretic Designs}

While PDL mitigation schemes have been either analyzed or
designed from an information-theoretic perspective in previous works
such as \cite{dumenil-dissertation,dumenil-jlt,dumenil-fundamental}, 
mutual information is not necessarily 
a proxy for \textit{practically} achievable rates
when the underlying channel is \textit{compound}. 
For example, consider a pair of parallel AWGN channels
with unknown SNR values $\mathsf{SNR}_1$ and $\mathsf{SNR}_2$
yet known constant sum-capacity $c$ so that 
\begin{equation}\label{constant-sum-capacity}
	\frac{1}{2}\log_2\mleft(1+\mathsf{SNR}_1\mright)
	+
	\frac{1}{2}\log_2\mleft(1+\mathsf{SNR}_2\mright) = c\text{.}
\end{equation}
Such a channel is referred to as a \textit{compound channel} \cite{compound-lapidoth}
with capacity $c$ since reliable communication at rate $c$ requires reliable
communication across \textit{every} parallel channel with 
arbitrary $\mathsf{SNR}_1$ and $\mathsf{SNR}_2$
satisfying \eqref{constant-sum-capacity}. 
One cannot expect an off-the-shelf coded modulation scheme  
designed for a scalar AWGN channel with an SNR of $\mathsf{SNR}_3$
such that
\begin{equation}\label{capacity-equivalent-channel}
	c = \frac{1}{2}\log_2\mleft(1+\mathsf{SNR}_3\mright)
\end{equation}
to achieve the same performance on \textit{every} parallel channel
satisfying \eqref{constant-sum-capacity} as it does on the capacity-equivalent scalar
channel. In fact, the problem of communication across a class 
of capacity-equivalent channels such as that described by \eqref{constant-sum-capacity} 
with a practical code
is an instance of the long-studied, non-trivial problem of the design of
\textit{universal codes} (see, e.g., \cite{maryam-universal,pramod-universal,tse-universal}). 
Therefore, invariance of mutual information across channel 
realizations under a certain PDL mitigation scheme does not
guarantee achievability by concatenation with 
a practical code designed for scalar AWGN channels
or simple modifications thereof.

Matters are further complicated when the parallel channels
are correlated as in the case of PDL-impaired PDM channels.
In such a situation, realization of an information-theoretic
promise could require high-complexity joint ML processing across the polarizations, 
and even then, we still have no performance guarantees under concatenation 
with codes designed for scalar channels. 

In contrast, the proposed scheme reduces the problem of 
communication across a PDL-impaired PDM channel
entirely to the classical, well-understood problem of scalar AWGN communication.
The scheme can thus be combined with standard practical 
coded modulation schemes designed for scalar AWGN channels such as those in \cite{bocherer-cookbook, masoud-cookbook} without loss of optimality. The
overall performance of the proposed scheme 
from a gap-to-capacity and frame error rate (FER) perspective
is then fully characterized in terms of the same measures on 
scalar AWGN channels for the constituent coded modulation schemes.

\section{Capacity of a PDL-Impaired Channel}\label{PDL-Cap}

Throughout this paper, we will use boldface font
for vectors and matrices with 
$(\cdot)_{ij}$ denoting the $ij$th entry of a matrix
and non-boldface font with subscripts denoting the
entries of a vector.

\subsection{Notions of Capacity and Compound Capacity}

We begin by considering the situation 
of a real-valued channel matrix. Without loss 
of generality, we can assume that the input
to the channel is real-valued with a complex 
input being interpreted as two uses of the real channel. 
In particular, we consider the two-parameter channel
defined by
\begin{equation*}
	\underbrace{\begin{bmatrix}
		Y_1 \\
		Y_2
	\end{bmatrix}}_{\mathbf{Y}}
	=
	\underbrace{\begin{bmatrix}
			\sqrt{1+\gamma} & 0 \\
			0 & \sqrt{1-\gamma}	
	\end{bmatrix}}_{\mathbf{D}_\gamma}
	\underbrace{\begin{bmatrix}
			\cos\theta & -\sin\theta \\
			\sin\theta & \cos\theta
	\end{bmatrix}}_{\mathbf{R}_\theta}
	\underbrace{\begin{bmatrix}
		X_1 \\
		X_2
	\end{bmatrix}}_{\mathbf{X}}
	+ 
	\underbrace{
	\begin{bmatrix}
		Z_1 \\
		Z_2
	\end{bmatrix}}_{\mathbf{Z}}
\end{equation*}
where  $\gamma \in [-\alpha,\alpha]$ and $\theta \in [0,2\pi)$
representing a class of channels with up to 
\begin{equation}\label{PDL}
	10\log_{10}\mleft(\frac{1+\alpha}{1-\alpha}\mright)\text{ dB}
\end{equation}
of PDL where $\alpha \in [0,1)$ is fixed.
Moreover, $\mathbf{X}$ and $\mathbf{Z}$ are independent
with $\mathbf{Z}$ being white Gaussian, denoted 
$\mathbf{Z}\sim \mathcal{N}(\mathbf{0},\mathbf{I}_2)$, 
and $\mathbf{X}$ satisfying the power constraint  
\begin{equation}\label{power-constraint}
\mathbb{E}\mleft[\norm{\mathbf X}_2^2\mright] 
\leq 2 \cdot \mathsf{SNR}\text{.}
\end{equation}
Note that while this model is inherently non-unitary
(or non-orthogonal), it is energy-preserving when 
$\mathbb{E}\mleft[X_1^2\mright] = \mathbb{E}\mleft[X_2^2\mright]$.
In particular, if $\mathbb{E}\mleft[X_1^2\mright] = \mathbb{E}\mleft[X_2^2\mright]$ 
and \eqref{power-constraint} holds with equality, then 
$\mathsf{SNR}$
is the ratio of the total received signal power to the total received noise power,
i.e., 
\begin{equation}\label{SNR-def}
	\mathsf{SNR}
	= 
	\frac{\mathbb{E}\mleft[\norm{\mathbf{D}_\gamma\mathbf{R}_\theta\mathbf{X}}_2^2\mright]}
	{\mathbb{E}\mleft[\norm{\mathbf{Z}}_2^2\mright]}
\end{equation}
as expected.

A variety of notions of channel capacity can be considered. One
notion is the classical Shannon capacity 
\begin{equation*}
	C_\mathsf{classical}(\gamma,\theta, \mathsf{SNR}) = 
	\max_{f_{\mathbf X},\,\mathbb{E}\mleft[\norm{\mathbf X}_2^2\mright] \leq 2 \cdot \mathsf{SNR}}
	I(\mathbf X;\mathbf{D}_\gamma\mathbf{R}_\theta\mathbf X+\mathbf Z)
\end{equation*}
where $f_{\mathbf X}$ denotes the probability density of $\mathbf X$ and 
$I(\cdot;\cdot)$ denotes mutual information. 
This represents the rate that can be achieved when 
$\gamma$ and $\theta$
are known at the transmitter. This is a well-understood problem 
\cite{Telatar-MIMO-Capacity} but does not represent our situation in which $\gamma$
and $\theta$ are not known at the transmitter since round-trip delays are typically
longer than the channel coherence time \cite{winzer-shtaif}. 

Another possibility is to assume some probability distribution over $\gamma$
and $\theta$ and define the \textit{ergodic} capacity
\begin{equation*}
	C_\mathsf{ergodic}(\mathsf{SNR}) = \mathbb{E}_{\gamma,\theta}\mleft[C_\mathsf{classical}(\gamma,\theta, \mathsf{SNR})\mright]\text{.}
\end{equation*}
This represents the rate achievable when the transmitted codeword
(or frame) spans many channel realizations, i.e., values of $\gamma$
and $\theta$. However, these parameters
are typically slowly varying relative to the baud rate 
so that achieving the ergodic capacity would require
averaging over a prohibitively large number of channel
uses \cite{dumenil-dissertation}.

The appropriate notion of capacity in this scenario is 
that of \textit{compound capacity} \cite{compound-lapidoth}. 
In particular, we define 
\begin{multline}\label{compound-capacity}
	C_\mathsf{compound}(\alpha,\mathsf{SNR}) = \\
	\max_{f_{\mathbf X},\,\mathbb{E}\mleft[\norm{\mathbf X}_2^2\mright] \leq 2\cdot\mathsf{SNR}}\;
	\min_{\gamma \in [-\alpha,\alpha],\,\theta \in [0,2\pi)}\,
	I(\mathbf X;\mathbf{D}_\gamma\mathbf{R}_\theta\mathbf X+\mathbf Z)\text{.}
\end{multline}
The compound capacity $C_\mathsf{compound}(\alpha,\mathsf{SNR})$ 
represents the rate of reliable communication 
that we can \textit{guarantee} assuming that 
$\gamma$ and $\theta$ 
are chosen by nature (or an adversary) from the sets $[-\alpha,\alpha]$ and $[0,2\pi)$, respectively, and then fixed for the duration of the transmission,
but are unknown to the transmitter which only knows $\alpha$ 
with the receiver knowing $\gamma$ and $\theta$.

Henceforth, we define the concise notation 
\begin{equation*}
	C_\alpha(\mathsf{SNR}) = C_\mathsf{compound}(\alpha,\mathsf{SNR})
\end{equation*}
and define the real scalar AWGN channel capacity function 
\begin{equation*}
	C(\mathsf{SNR}) = \frac{1}{2}\log_2(1+\mathsf{SNR})
\end{equation*}
and proceed to compute \eqref{compound-capacity}. 
A standard information-theoretic argument 
as in \cite{Telatar-MIMO-Capacity} will show that we
can omit $\mathbf{R}_\theta$ from the calculation 
and take $X_1$ and $X_2$ to be independent with
\begin{align*}
	X_1 &\sim \mathcal N(0, 2\beta \mathsf{SNR})\text{,} \\
	X_2 &\sim \mathcal N(0, 2(1-\beta) \mathsf{SNR})\text{,} 
\end{align*}
and $\beta\in[0,1]$ a power allocation factor. The capacity
calculation then reduces to 
\begin{multline*}\label{compound-reduction}
	C_\alpha(\mathsf{SNR}) 
	= \\ 
	\max_{\beta\in[0,1]}\,
	\min_{\gamma\in[-\alpha,\alpha]}\,
	C(2(1+\gamma) \beta \mathsf{SNR}) + 
	C(2(1-\gamma) (1-\beta) \mathsf{SNR})\text{.}
\end{multline*}

We begin with the inner minimization. 
Define $\gamma^*(\beta)$ by
\begin{multline*}
	\gamma^*(\beta) 
	= \\	\argmin_{\gamma\in[-\alpha,\alpha]}
	C(2(1+\gamma) \beta \mathsf{SNR}) + 
	C(2(1-\gamma) (1-\beta) \mathsf{SNR})\text{.}
\end{multline*}
For a fixed $\beta$, a simple convexity 
argument will show that 
\begin{equation*}
	\gamma^*(\beta)=
	\begin{cases}
		\alpha & \text{if } \beta \leq 1/2\\
		-\alpha & \text{if } \beta  > 1/2\\
	\end{cases}\text{,}
\end{equation*}
i.e., that the worst-case PDL is always extremal.
Proceeding to the outer 
maximization, define $\beta^*$ by
\begin{multline*}
	\beta^*=\\
	\argmax_{\beta\in[0,1]}
	C(2(1+\gamma^*(\beta)) \beta \mathsf{SNR}) + 
	C(2(1-\gamma^*(\beta)) (1-\beta) \mathsf{SNR})\text{.}
\end{multline*}
Elementary algebra shows that $\beta^* = 1/2$, i.e.,
that the optimal power allocation is symmetric, and we get that
\begin{equation}\label{PDL-capacity}
C_\alpha(\mathsf{SNR}) =
C((1+\alpha)\mathsf{SNR}) + 
C((1-\alpha)\mathsf{SNR})\text{.}
\end{equation}

Thus, we see that the compound capacity
of a PDL-impaired channel \eqref{PDL-capacity} 
corresponds to the sum-capacity of the two
polarizations when $\theta=0$, 
$\gamma = \pm \alpha$, and $\mathbb{E}\mleft[X_1^2\mright] = \mathbb{E}\mleft[X_2^2\mright] = \mathsf{SNR}$.
While this result and analysis is implicit in previous works such 
as \cite{dumenil-dissertation,dumenil-fundamental}, the framework
of compound capacity formalizes the problem and allows us
to reason more carefully about it.

Finally, we remark that compound capacity is essentially a proxy for \textit{outage capacity} \cite{foschini} when the channel parameters are random.
In particular, the probability of outage is determined by the PDL distribution; a fixed outage probability translates to a bound on the PDL, i.e., a value for $\alpha$.
A typical value is $\alpha = 0.599$ corresponding to $6$ dB of PDL and an outage 
probability of $10^{-5}$ \cite{dumenil-dissertation}. 
However, as noted earlier, a capacity-achieving scheme under the compound
channel model considered in this paper is \textit{not} 
outage optimal if we wish to additionally model uncertainty in the 
polarization-average loss or IL as in \cite{foschini}.

\subsection{Capacity Under Non-Joint Coding}

\begin{figure}[t]
	\centering
	\includegraphics[width=\columnwidth]{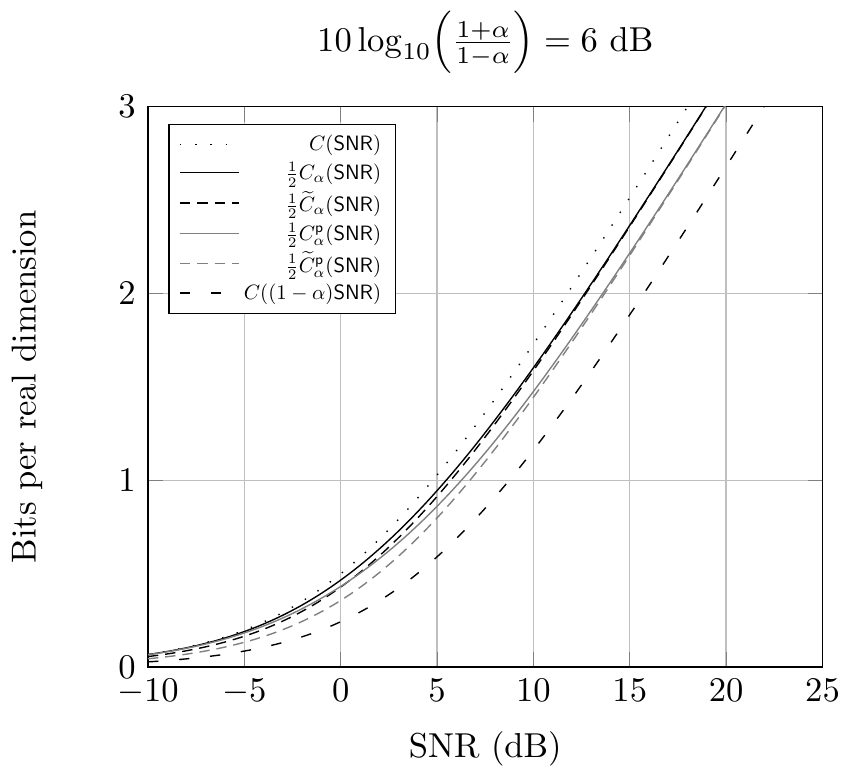}
	\caption{Compound channel capacities, asymptotic approximations, and AWGN capacity all normalized by the number of real dimensions.}\label{Capacities-Figure}
\end{figure}

To simplify the exposition, we will work with
high-SNR approximations and consider the use of zero-forcing
(ZF) instead of LMMSE receivers first. However, we emphasize that
every approximation is a rigorous asymptotic and is accompanied by 
a corresponding exact result.
We define a high-SNR approximation to the
capacity \eqref{compound-capacity}, denoted $\widetilde C_\alpha(\mathsf{SNR})$,
by 
\begin{equation}\label{compound-capacity-approx}
	\widetilde C_\alpha(\mathsf{SNR}) = C((1-\alpha^2)\mathsf{SNR}) 
	+ C(\mathsf{SNR})\text{.}
\end{equation}
It can be shown that  
\begin{equation*}
	C_\alpha(\mathsf{SNR}) = \widetilde C_\alpha(\mathsf{SNR}) + \mathcal{O}(\mathsf{SNR}^{-1})
\end{equation*}
where the hidden constant depends on $\alpha$.

\begin{figure}[t]
	\centering
	\includegraphics[width=\columnwidth]{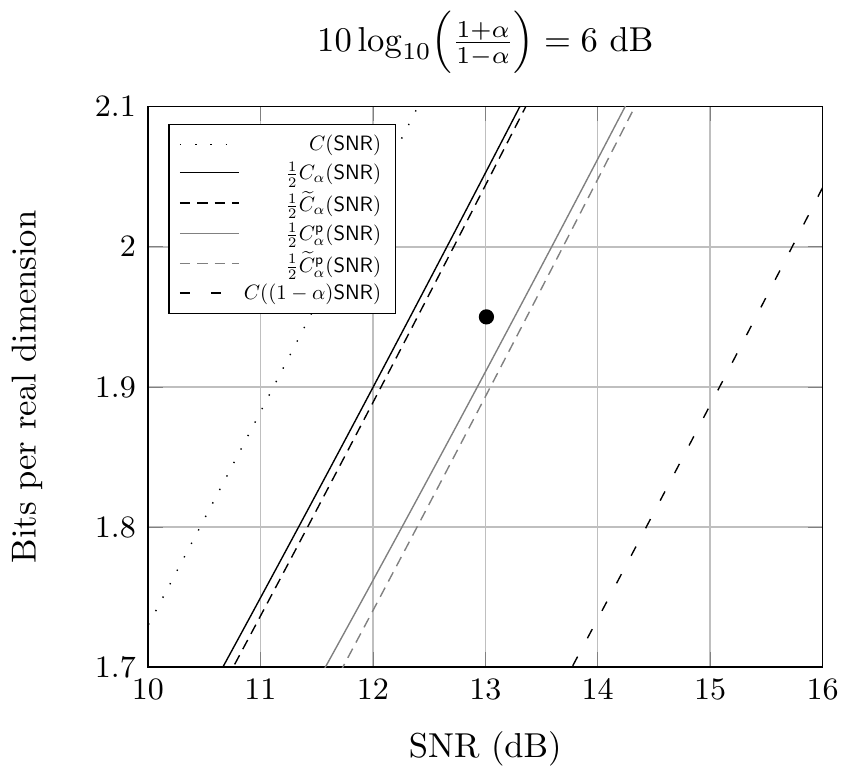}
	\caption{Close-up of Fig.~\ref{Capacities-Figure}. The solid circle is an operating
	point achievable by use of the proposed scheme in conjunction with the practical coded modulation scheme from \cite{bocherer-cookbook} described in Section \ref{Concrete}.}\label{Capacities-Figure-Zoomed}
\end{figure}

We will now establish the achievable rate under non-joint 
coding in which separate codewords are transmitted on each
polarization and decoded separately. 
Henceforth, we will take $\mathbf{X} \sim \mathcal{N}(\mathbf{0}, \mathsf{SNR}\mathbf{I}_2)$
so that 
\begin{equation}\label{MI}
	I(X_1 X_2; Y_1 Y_2) 
	= 
	C((1+\gamma)\mathsf{SNR}) + 
	C((1-\gamma)\mathsf{SNR})
\end{equation}
and 
\begin{equation*}
	\min_{\gamma,\theta}I(X_1 X_2; Y_1 Y_2)
	= C_\alpha(\mathsf{SNR})
\end{equation*}
where the minimization is over $\gamma \in [-\alpha,\alpha]$ 
and $\theta \in [0,2\pi)$ with these sets suppressed for brevity.
Note that by the chain rule of mutual information \textit{and} independence of $X_1$ and $X_2$,
we have (see, e.g., \cite[Chapter~8]{Cover-Thomas}) that
\begin{equation}\label{chain-rule}
	I(X_1 X_2; Y_1 Y_2) = I(X_1; Y_1Y_2) + I(X_2; Y_1Y_2X_1)\text{.}
\end{equation}
Moreover, one can compute (see, e.g., \cite[Chapter~9]{Cover-Thomas}) that
\begin{multline*}
	I(X_1; Y_1 Y_2) = \\
	C((1+\gamma)\mathsf{SNR})
	+ 
	C((1-\gamma)\mathsf{SNR})
	-
	C((1-\gamma\cos(2\theta))\mathsf{SNR})
\end{multline*}
and that
\begin{equation*}
	I(X_2; Y_1 Y_2 X_1)
	=
	C((1-\gamma\cos(2\theta))\mathsf{SNR})
\end{equation*}
which add up to \eqref{MI} as one expects
from the chain rule of mutual information
\eqref{chain-rule}.
Note that these mutual information terms
in the chain rule of mutual information
are precisely the capacities of the 
sub-channels induced by LMMSE-SIC 
when we have an AWGN MIMO channel  (see, e.g., \cite[Chapter~8]{tse-wireless-communication}
and \cite{elaine}).

We then see that
\begin{align*}
	C_\alpha(\mathsf{SNR}) &= \min_{\gamma,\theta}I(X_1 X_2; Y_1 Y_2) \\
	&= \min_{\gamma,\theta} \big[I(X_1; Y_1Y_2) + I(X_2; Y_1Y_2X_1)\big]\\
	&\geq
	\min_{\gamma,\theta} I(X_1; Y_1Y_2) + \min_{\gamma,\theta} I(X_2; Y_1Y_2X_1)\\
	&= 2C((1-\alpha)\mathsf{SNR}) 
\end{align*}
which highlights the fundamental problem of communication
across this channel. This problem is the mismatch between
the minimum of
the sum and the sum of the minima of the capacities
of the two polarizations. Therefore, if each polarization
is treated as a separate channel, we can only guarantee a
rate of $C((1-\alpha)\mathsf{SNR})$ on each even if we perfectly
remove the interference from one of them.

Note, however, that this achievable rate is a highly pessimistic
baseline since merely interleaving a bit stream across the two polarizations
constitutes a form of joint coding and could result in a smaller penalty
than \eqref{nj-penalty}.

\subsection{Parallel Capacity}

We now define the capacity achievable
under a parallel decoding architecture whereby 
no interference cancellation or decision feedback 
across the polarizations is allowed.
Noting that 
\begin{equation*}
	I(X_1 X_2; Y_1 Y_2) \geq I(X_1; Y_1Y_2) + I(X_2; Y_1Y_2)\text{,}
\end{equation*}
and that
\begin{multline*}
	I(X_2; Y_1 Y_2)
	=\\
	C((1+\gamma)\mathsf{SNR})
	+ 
	C((1-\gamma)\mathsf{SNR})
	-
	C((1+\gamma\cos(2\theta))\mathsf{SNR})\text{,}
\end{multline*}
we define and compute the parallel
capacity $C_\alpha^\mathsf{p}(\mathsf{SNR})$ as
		\begin{align*}
	C_\alpha^\mathsf{p}(\mathsf{SNR}) 
	&= 
	\min_{\gamma,\theta} \big[I(X_1; Y_1Y_2) + I(X_2; Y_1Y_2)\big]\\
	&= 2C_\alpha(\mathsf{SNR})-2C(\mathsf{SNR})\\
	&= 
	2\widetilde C_\alpha(\mathsf{SNR})-2C(\mathsf{SNR}) + 
	\mathcal{O}(\mathsf{SNR}^{-1})\\
	&= 
	2C((1-\alpha^2)\mathsf{SNR}) + \mathcal{O}(\mathsf{SNR}^{-1})\text{,}
\end{align*}
where we have substituted the approximation \eqref{compound-capacity-approx}.
We accordingly define a high-SNR approximation for the
parallel capacity $\widetilde C_\alpha^\mathsf{p}(\mathsf{SNR})$
by 
\begin{equation*}
	\widetilde C_\alpha^\mathsf{p}(\mathsf{SNR})
	= 
	2C((1-\alpha^2)\mathsf{SNR}) \text{.}
\end{equation*}

These results are plotted in Fig.~\ref{Capacities-Figure}
and Fig.~\ref{Capacities-Figure-Zoomed} for $\alpha=0.599$
which corresponds to a $6$ dB worst-case PDL. One can see from
the figures penalties relative to a PDL-free channel of 
roughly $1$ dB under an optimal scheme, $2$ dB under a 
parallel scheme, and $4$ dB under a non-joint scheme.
Alternatively, these penalties at high SNR can be 
computed by \eqref{nj-penalty}, \eqref{p-penalty}, and
\eqref{sic-penalty}. Moreover, we remark that in the limit
of high PDL, i.e., $\alpha$ approaching $1$, the SNR
penalties become infinite: not because the capacity approaches zero,
but because its growth rate is halved relative to the reference
PDL-free channel.

\subsection{AWGN Channels Induced by Linear Equalization}\label{SNR-Calculation-Section}

Before proceeding to describe our capacity-achieving schemes,
we will review---for the reader's convenience---the method of calculating SNRs and noise statistics 
under linear precoding and equalization with possible interference cancellation. 
Moreover, note 
that for the remainder of this paper, we may redefine 
previously used notation where appropriate. 

The effective channel after precoding and possibly interference
cancellation is given by 
\begin{equation*}
	\mathbf{Y} = \mathbf{H}\mathbf{U} + \mathbf{Z}
\end{equation*}
where $\mathbf{H}\in \mathbb{R}^{m \times n}$ with $m \geq n$, 
\begin{align*}
	\mathbb{E}\mleft[\mathbf{U}\mathbf{U}^\intercal\mright] &= \mathsf{SNR}\mathbf{I}_n\text{,}\\
	\mathbb{E}\mleft[\mathbf{U}\mright] &= \mathbf{0}\text{,}
\end{align*}
and $\mathbf{Z}\sim \mathcal{N}(\mathbf{0}, \mathbf{I}_m)$ with
$\mathbf{U}$ and $\mathbf{Z}$ independent so that 
\begin{equation}\label{zero-cross}
	\mathbb{E}\mleft[\mathbf{U}\mathbf{Z}^\intercal\mright] = \mathbf{0}_{n\times m}\text{.}
\end{equation}
After multiplication by an equalization matrix $\mathbf{E} \in \mathbb{R}^{n \times m}$, the effective channel matrix becomes $\mathbf{E}\mathbf{H} \in \mathbb{R}^{n \times n}$. 
We additively decompose $\mathbf{E}\mathbf{H}$ as 
\begin{equation*}
	\mathbf{EH} = \mathbf{\Lambda} + \mathbf{F}
\end{equation*}
where $\mathbf{\Lambda}\in\mathbb{R}^{n \times n}$ is a diagonal matrix whose diagonal entries
are the diagonal entries of $\mathbf{E}\mathbf{H}$ so 
that $\mathbf{F} = \mathbf{E}\mathbf{H} - \mathbf{\Lambda}$
is a matrix with zeros on the diagonal.

The effective channel after equalization is
then given by 
\begin{align*}
	\widetilde{\mathbf Y}
	= 
	\mathbf{E}\mathbf{Y}
	&= 
	\mathbf{E}\mathbf{H}\mathbf{U} + \mathbf{E}\mathbf{Z}\\
	&=
	\underbrace{\mathbf{\Lambda}\mathbf{U}}_{\widetilde{\mathbf{U}}}
	+ 
	\underbrace{\mathbf{F}\mathbf{U} + \mathbf{E}\mathbf{Z}}_{\widetilde{\mathbf{Z}}}\\
	&=
	\mathbf{\widetilde{U}} + \mathbf{\widetilde{Z}}
\end{align*}
where $\widetilde{\mathbf{U}}$ is our new scaled signal vector
and $\mathbf{\widetilde{Z}}$ is our new noise vector which includes
interference noise.

We then have covariance and cross-covariance matrices
given by 
\begin{align}
	\mathbf{K}_{\widetilde{\mathbf{U}}\widetilde{\mathbf{U}}}
	&= \mathbb{E}\mleft[\widetilde{\mathbf{U}}\widetilde{\mathbf{U}}^\intercal\mright]	
	= \mathsf{SNR}\mathbf{\Lambda}\mathbf{\Lambda}^\intercal 
	= \mathsf{SNR}\mathbf{\Lambda}^2\\
\mathbf{K}_{\widetilde{\mathbf{U}}\widetilde{\mathbf{Z}}}
&= \mathbb{E}\mleft[\widetilde{\mathbf{U}}\widetilde{\mathbf{Z}}^\intercal\mright]
= \mathsf{SNR}\mathbf{\Lambda}\mathbf{F}^\intercal\\
\mathbf{K}_{\widetilde{\mathbf{Z}}\widetilde{\mathbf{Z}}}
&= \mathbb{E}\mleft[\widetilde{\mathbf{Z}}\widetilde{\mathbf{Z}}^\intercal\mright]	
= \mathsf{SNR}\mathbf{F}\mathbf{F}^\intercal + \mathbf{E}\mathbf{E}^\intercal
\end{align}
where we have used \eqref{zero-cross}.

Note that, by design, we have
\begin{equation}
	\mathbb{E}[\widetilde{U}_i\widetilde{Z}_i]
	=
	\left(\mathbf{K}_{\widetilde{\mathbf{U}}\widetilde{\mathbf{Z}}}\right)_{ii} = 0\text{.}
\end{equation}
for $i \in\{1,2,\dots, n\}$. 
Therefore, we can define SNRs for
our additive noise sub-channels by 
\begin{equation}
	\mathsf{SNR}_i = 
	\frac{\mathbb{E}[\widetilde{U}_i^2]}{\mathbb{E}[\widetilde{Z}_i^2]}
	= 
	\frac{\left(\mathbf{K}_{\widetilde{\mathbf{U}}\widetilde{\mathbf{U}}}\right)_{ii}}
	{\left(\mathbf{K}_{\widetilde{\mathbf{Z}}\widetilde{\mathbf{Z}}}\right)_{ii}}
\end{equation}
for $i \in\{1,2,\dots, n\}$.

In the discussions which follow, we will assume that the input 
signal is Gaussian, i.e., 
$\mathbf{U}\sim \mathcal{N}(\mathbf{0}, \mathsf{SNR}\mathbf{I}_n)$, 
in which case the interference noise is Gaussian and these sub-channels 
are strictly AWGN channels. Despite the fact that, in practice, $\mathbf{U}$
will contain discrete entries from, say, a PAM constellation so that
$\mathbf{\widetilde Z}$ is not strictly Gaussian, this
assumption will not be of any material significance for two reasons.
Firstly, it
is known from information theory that we can safely
approximate additive noise as Gaussian 
and guarantee AWGN rates under AWGN decoding metrics (see, e.g., \cite[Chapter~9]{Cover-Thomas}).
Secondly, 
we will nonetheless provide a ZF-SIC scheme which achieves \eqref{compound-capacity-approx}
and thus \eqref{sic-penalty}  
in which case $\mathbf{F} = \mathbf{0}_{n\times n}$,
$\widetilde{\mathbf{Z}}$ is Gaussian, and we are 
guaranteed a synthesis of strictly AWGN sub-channels
without any such assumption.

At this point, we see that under separate coding, i.e., 
transmission of separate codewords across each sub-channel,
we can achieve a rate of $\sum_{i=1}^{n} C(\mathsf{SNR}_i)$.
Supposing now that $\mathsf{SNR}_j = \mathsf{SNR}_k$ for some 
$j, k\in\{1,2,\dots, n\}$ with $j\neq k$, one might be tempted to treat
the $j$th and $k$th sub-channels as a single AWGN channel 
and spread one codeword across them.
However, the codeword will \textit{not} see a classical AWGN channel
unless we further have that 
\begin{align*}
	\mathbb{E}[\widetilde{U}_j\widetilde{Z}_k]
	&=
	\left(\mathbf{K}_{\widetilde{\mathbf{U}}\widetilde{\mathbf{Z}}}\right)_{jk} = 0\\
	\mathbb{E}[\widetilde{Z}_j\widetilde{Z}_k]
	&=
	\left(\mathbf{K}_{\widetilde{\mathbf{Z}}\widetilde{\mathbf{Z}}}\right)_{jk} = 0\text{,}
\end{align*}
i.e., no noise--noise
or signal--noise correlations across the codeword. While 
this may or may not be an issue in practice, 
we cannot claim a performance guarantee in terms of the classical
AWGN performance of the constituent code unless the code sees a statistically equivalent channel. This is
only guaranteed by sending different codewords on each channel.

We now provide an example demonstrating
the achievable rate under ZF with no joint coding. 
Taking $\mathbf{X} = \mathbf{U}$, i.e., no precoding, we have
a channel matrix of $\mathbf{H} = \mathbf{D}_\gamma\mathbf{R}_\theta$
and ZF equalizer given by $\mathbf{E}=\mathbf{H}^{-1} = \mathbf{R}_\theta^\intercal\mathbf{D}_\gamma^{-1}$. This yields
\begin{align*}
	\mathbf{K}_{\widetilde{\mathbf{U}}\widetilde{\mathbf{U}}} 
	&= \mathsf{SNR}\mathbf{I}_2\\
	\mathbf{K}_{\widetilde{\mathbf{U}}\widetilde{\mathbf{Z}}} 
	&= \mathbf{0}_{2\times 2}
\end{align*}
as expected and that
\begin{equation*}
	\mathbf{K}_{\widetilde{\mathbf{Z}}\widetilde{\mathbf{Z}}} 
	=
	\begin{bmatrix}
		\frac{1-\gamma\cos(2\theta)}{1-\gamma^2} & \frac{\gamma\sin(2\theta)}{1-\gamma^2}\\
		\frac{\gamma\sin(2\theta)}{1-\gamma^2} & \frac{1+\gamma\cos(2\theta)}{1-\gamma^2} 
	\end{bmatrix}\text{.}
\end{equation*}
We then have $\gamma$- and $\theta$-dependent SNRs of 
\begin{align*}
	\mathsf{SNR}_1(\gamma,\theta) &=
	\frac{(1-\gamma^2)\mathsf{SNR}}{1-\gamma\cos(2\theta)}\\
	&\geq 
	\frac{(1-\alpha^2)\mathsf{SNR}}{1+\alpha}\\
	&=(1-\alpha)\mathsf{SNR}
\end{align*}
and
\begin{align*}
	\mathsf{SNR}_2(\gamma,\theta) &=
	\frac{(1-\gamma^2)\mathsf{SNR}}{1+\gamma\cos(2\theta)}\\
	&\geq 
	\frac{(1-\alpha^2)\mathsf{SNR}}{1+\alpha}\\
	&=(1-\alpha)\mathsf{SNR}\text{.}
\end{align*}
Thus, we have a capacity of $C((1-\alpha)\mathsf{SNR})$
for each sub-channel. Somewhat counter-intuitively,
this shows that in the absence of joint coding across
the two polarizations, one can do no better 
than ZF. To better understand this, simply note that 
ZF is optimal for $\theta = 0$ and any gains obtained
by LMMSE or LMMSE-SIC for different values of $\theta$
are irrelevant because it is only the worst case that matters.

\section{Optimal Scheme for Real-Valued PDL Channels}\label{Real-Scheme}

\subsection{Setup}

We now proceed to provide a capacity-achieving scheme
for the case of a real-valued channel. Consider a
two-channel-use extension of our channel so
that the channel matrix is 
\begin{equation*}
	\diag\mleft(\mathbf{D}_\gamma\mathbf{R}_\theta, \mathbf{D}_\gamma\mathbf{R}_\theta \mright)\in 
	\mathbb{R}^{4\times 4}\text{.}
\end{equation*}
Consider then taking the input to the channel to
be $\mathbf{X} = \mathbf{G}\mathbf{U}$ where $\mathbf{G} \in \mathbb{R}^{4\times 4}$
is an orthogonal precoding matrix so that $\mathbf{\mathbf{G}\mathbf{G}^\intercal} = \mathbf{G}^\intercal\mathbf{G} = \mathbf{I}_4$ and 
$\mathbb{E}[\mathbf{X}\mathbf{X}^\intercal] =
\mathsf{SNR}\mathbf{I}_4$ when $\mathbb{E}[\mathbf{U}\mathbf{U}^\intercal] =
\mathsf{SNR}\mathbf{I}_4$.
The effective channel matrix is then 
	$$\mathbf{H} = \diag\mleft(\mathbf{D}_\gamma\mathbf{R}_\theta, \mathbf{D}_\gamma\mathbf{R}_\theta \mright)\mathbf{G}$$
with the effective channel being 
$\mathbf{Y} = \mathbf{H}\mathbf{U} + \mathbf{Z}$
where $\mathbf{Z}\sim\mathcal{N}(\mathbf{0},\mathbf{I}_4)$ 
so that when $\mathbf{U}\sim \mathcal{N}(\mathbf{0},\mathsf{SNR}\mathbf{I}_4)$,
we have
\begin{align*}
	2C_\alpha(\mathsf{SNR})
	&= 
	\min_{\gamma,\theta}
	I(\mathbf{U};\mathbf{Y})\\
	&=
	\min_{\gamma,\theta} 
	\bigg[ 
	I(U_1;\mathbf{Y}) 
	+ I(U_2;\mathbf{Y} U_1) \\
	&\phantom{=\min_{\gamma,\theta}}+ I(U_3;\mathbf{Y} U_1 U_2)  +
	I(U_4;\mathbf{Y} U_1 U_2 U_3)\bigg] \text{.}
\end{align*}

To achieve capacity under LMMSE-SIC, it suffices
to find a precoder $\mathbf{G}$ such that 
\begin{align*}
	&\min_{\gamma,\theta} 
	\bigg[ 
	I(U_1;\mathbf{Y}) 
	+ I(U_2;\mathbf{Y} U_1) \\
	&\phantom{\min_{\gamma,\theta}}+ I(U_3;\mathbf{Y} U_1 U_2)  +
	I(U_4;\mathbf{Y} U_1 U_2 U_3)\bigg]\\
	&=
	\min_{\gamma,\theta} 	I(U_1;\mathbf{Y}) 
	+ \min_{\gamma,\theta} I(U_2;\mathbf{Y}U_1)\\
	&\phantom{=}+
	\min_{\gamma,\theta} I(U_3;\mathbf{Y} U_1 U_2) +
	\min_{\gamma,\theta} I(U_4;\mathbf{Y} U_1 U_2 U_3)\text{.}
	\tag{$\star$}
\end{align*}
Surprisingly, such a precoder exists and is given by
\begin{equation}\label{magic-1}
	\mathbf{G}
	= 
	\frac{1}{\sqrt{2}}
	\begin{bmatrix}
		1 & 0 & 1 & 0\\
		0 & 1 & 0 & 1\\
		0 & 1 & 0 & -1\\
		-1 & 0 & 1 & 0
	\end{bmatrix}\text{.}
\end{equation}
Importantly, even a minor variation on this
precoder such as that in \cite{huawei} will not
satisfy ($\star$). This is due to the fact that
while the
left-hand side of ($\star$) is invariant
under permutations of $U_1,U_2,U_3,U_4$, or 
equivalently, column permutations of $\mathbf{G}$,
the right-hand side is not. 

The existence of this precoder is surprising because, 
for a general compound MIMO channel, a precoding matrix which makes ($\star$) true or approximately true must generally be a function of the channel matrix (in our case, the parameters $\gamma$ and $\theta$) thus resulting in a scheme which requires channel knowledge at the transmitter.
We are nonetheless able to find a channel-independent precoding matrix \eqref{magic-1} which satisfies ($\star$) because of the mathematical peculiarities of the particular class of channels under consideration. We elaborate on this point in the Appendix.

We will proceed to demonstrate that \eqref{magic-1},
in fact, 
satisfies the stronger property that 
\begin{multline*}
	\min_{\gamma,\theta}I(U_1;\mathbf{Y}) 
	+ 
	\min_{\gamma,\theta}I(U_2;\mathbf{Y}) \\
	+
	\min_{\gamma,\theta}I(U_3;\mathbf{Y}U_1U_2) 
	+ 
	\min_{\gamma,\theta}I(U_4;\mathbf{Y}U_1U_2) \\
	= 2C_\alpha(\mathsf{SNR})\text{.}
\end{multline*}
However, we will first demonstrate 
that ZF-SIC, which is much simpler to analyze, achieves the 
high-SNR approximation to the capacity $2\widetilde C_\alpha(\mathsf{SNR})$
under the precoding \eqref{magic-1}.

\subsection{ZF and ZF-SIC}\label{real-ZF-SIC}

With $\mathbf{G}$ as in \eqref{magic-1}, we get the 
effective channel matrix 
\begin{multline*}
	\mathbf{H} = \frac{1}{\sqrt{2}}
	\diag\mleft(\sqrt{1+\gamma},\sqrt{1-\gamma},\sqrt{1+\gamma},\sqrt{1-\gamma}\mright)\\
	\cdot
	\begin{bmatrix}
		\cos(\theta) & -\sin(\theta) & \cos(\theta) & -\sin(\theta)\\
		\sin(\theta) & \cos(\theta) & \sin(\theta) & \cos(\theta)\\
		\sin(\theta) & \cos(\theta) & -\sin(\theta) & -\cos(\theta)\\
		-\cos(\theta) & \sin(\theta) & \cos(\theta) & -\sin(\theta)
	\end{bmatrix}\text{.}
\end{multline*}
By taking $\mathbf{E} = \mathbf{H}^{-1}$, i.e., the ZF equalizer, and 
proceeding as in Section \ref{SNR-Calculation-Section},
one finds that
\begin{align*}
	\mathbf{K}_{\widetilde{\mathbf{U}}\widetilde{\mathbf{U}}} 
	&= \mathsf{SNR}\mathbf{I}_4\\
	\mathbf{K}_{\widetilde{\mathbf{U}}\widetilde{\mathbf{Z}}} 
	&= \mathbf{0}_{4\times 4}
\end{align*}
as expected and that
\begin{equation*}
	\mathbf{K}_{\widetilde{\mathbf{Z}}\widetilde{\mathbf{Z}}} 
	=
	\begin{bmatrix}
		\frac{1}{1-\gamma^2} & 0 & \frac{-\gamma\cos(2\theta)}{1-\gamma^2} & \frac{\gamma\sin(2\theta)}{1-\gamma^2}\\
		0 & \frac{1}{1-\gamma^2} & \frac{\gamma\sin(2\theta)}{1-\gamma^2} & \frac{\gamma\cos(2\theta)}{1-\gamma^2}\\
		\frac{-\gamma\cos(2\theta)}{1-\gamma^2} & \frac{\gamma\sin(2\theta)}{1-\gamma^2} & \frac{1}{1-\gamma^2} & 0\\
		\frac{\gamma\sin(2\theta)}{1-\gamma^2} & \frac{\gamma\cos(2\theta)}{1-\gamma^2} & 0 &  \frac{1}{1-\gamma^2}
	\end{bmatrix}\text{.}
\end{equation*}
We then have for $i \in \{1,2,3,4\}$
that 
\begin{align*}
	\mathsf{SNR}_i(\gamma,\theta) &= (1-\gamma^2)\mathsf{SNR} \\ 
	&\geq (1-\alpha^2)\mathsf{SNR}\text{.}
\end{align*}
We can then achieve a rate of $C((1-\alpha^2)\mathsf{SNR})$ on each sub-channel  
and can thus achieve the high-SNR approximation to the parallel capacity 
$$2\widetilde C_\alpha^\mathsf{p}(\mathsf{SNR}) = 4C((1-\alpha^2)\mathsf{SNR})\text{.}$$
Note that this is also accomplished by schemes provided in \cite{oyama,pairwise} and we
will consider proceeding differently instead.

Consider coding across many channel uses of the first and second
sub-channels, which are classical AWGN channels whose SNRs satisfy
\begin{align*}
	\mathsf{SNR}_1(\gamma,\theta) &\geq (1-\alpha^2)\mathsf{SNR}\\
	\mathsf{SNR}_2(\gamma,\theta) &\geq (1-\alpha^2)\mathsf{SNR}\text{.}
\end{align*}
We can then recover $U_1$ and $U_2$ with arbitrarily high reliability
by either using a sufficiently strong coded modulation scheme or
backing away from capacity as with any AWGN channel.
We then assume that $U_1$ and $U_2$ have been decoded correctly
and cancel the corresponding interference.

One can verify that we have
\begin{multline*}
	\mathbf{H}^\intercal\mathbf{H} = \\
	\begin{bmatrix}
		1 & 0 & \gamma\cos(2\theta) & -\gamma\sin(2\theta)\\
		0 & 1 & -\gamma\sin(2\theta) & -\gamma\cos(2\theta) \\
		\gamma\cos(2\theta) & -\gamma\sin(2\theta) & 1 & 0\\
		-\gamma\sin(2\theta) & -\gamma\cos(2\theta) & 0 & 1
	\end{bmatrix}
\end{multline*}
so that we can partition $\mathbf{H}$ as
\begin{equation*}
	\mathbf{H} = 
	\begin{bmatrix}
		\mathbf{H}_1 & \mathbf{H}_2
	\end{bmatrix}
\end{equation*}
where $\mathbf{H}_1,\mathbf{H}_2 \in \mathbb{R}^{4\times 2}$ and satisfy
$\mathbf{H}_1^\intercal\mathbf{H}_1 = \mathbf{H}_2^\intercal\mathbf{H}_2 = \mathbf{I}_2$.

Cancelling the interference from $U_1$ and $U_2$ then yields
the effective channel
\begin{equation*}
	\mathbf{Y}-\mathbf{H}_1\begin{bmatrix}
		U_1 \\
		U_2
	\end{bmatrix} 
	= \mathbf{H}_2\begin{bmatrix}
		U_3 \\
		U_4
	\end{bmatrix} 
	+
	\mathbf{Z}
\end{equation*}
which we can optimally equalize with $\mathbf{E}=\mathbf{H}_2^\intercal$ to get
\begin{equation*}
	\mathbf{\widehat{Y}} = 
	\mathbf{H}_2^\intercal\left(\mathbf{Y}-\mathbf{H}_1\begin{bmatrix}
		U_1 \\
		U_2
	\end{bmatrix}\right) 
	= \underbrace{\begin{bmatrix}
			U_3 \\
			U_4
	\end{bmatrix}}_{\mathbf{\widehat{U}}} + \underbrace{\mathbf{H}_2^\intercal\mathbf{Z}}_{\mathbf{\widehat{Z}}}
\end{equation*}
which yields
\begin{align*}
	\mathbf{K}_{\widehat{\mathbf{U}}\widehat{\mathbf{U}}} 
	&= \mathsf{SNR}\mathbf{I}_2\\
	\mathbf{K}_{\widehat{\mathbf{U}}\widehat{\mathbf{Z}}} 
	&= \mathbf{0}_{2\times 2}\\
	\mathbf{K}_{\widehat{\mathbf{Z}}\widehat{\mathbf{Z}}} 
	&= \mathbf{I}_{2}\text{.}
\end{align*}

The SNRs seen by $U_1,U_2,U_3,U_4$ are then given by
\begin{align*}
	\mathsf{SNR}_1(\gamma,\theta) &\geq (1-\alpha^2)\mathsf{SNR}\\
	\mathsf{SNR}_2(\gamma,\theta) &\geq (1-\alpha^2)\mathsf{SNR}\\
	\mathsf{SNR}_3(\gamma,\theta) &= \mathsf{SNR}\\
	\mathsf{SNR}_4(\gamma,\theta) &= \mathsf{SNR}
\end{align*}
and we can thus achieve, with the ZF-SIC scheme and the precoder \eqref{magic-1}, 
$$2\widetilde C_\alpha(\mathsf{SNR}) = 2C((1-\alpha^2)\mathsf{SNR}) + 2C(\mathsf{SNR})$$
which is within $\mathcal{O}(\mathsf{SNR}^{-1})$ of the true capacity $2C_\alpha(\mathsf{SNR})$. 

Moreover, we have
no correlations within the first and second sub-channels, as well 
as within the third and fourth sub-channels. As a result, we can can combine
them and only need to send two codewords from two different codes having
two different rates. One code will see an SNR of $(1-\alpha^2)\mathsf{SNR}$
at worst, and the other will see an SNR of $\mathsf{SNR}$.

\subsection{LMMSE and LMMSE-SIC}

We now demonstrate how to fully close the gap between $2\widetilde C_\alpha(\mathsf{SNR})$
and $2C_\alpha(\mathsf{SNR})$ even though it is quite negligible as can be seen in Fig.~\ref{Capacities-Figure-Zoomed}.
This is merely a matter of replacing ZF in the scheme just described with LMMSE. While the calculations will become
somewhat tedious moving forward, the results will be essentially the same. 
By taking 
\begin{equation*}
	\mathbf{E} =
	\mathbf{H}^\intercal\left(\mathbf{H}\mathbf{H}^\intercal + \frac{1}{\mathsf{SNR}}\mathbf{I}_4\right)^{-1}
\end{equation*}
and proceeding as in Section \ref{SNR-Calculation-Section}, 
one obtains 
\begin{equation*}
	\mathbf{K}_{\mathbf{\widetilde{U}}\mathbf{\widetilde{U}}}
	=
	\frac{\mathsf{SNR}^3(\mathsf{SNR}(1-\gamma^2) + 1)^2}{(\mathsf{SNR}^2(1-\gamma^2) + 2\mathsf{SNR} + 1)^2}
	\mathbf{I}_4\text{.}
\end{equation*}
Moreover, we get
\begin{equation*}
	\mathbf{K}_{\mathbf{\widetilde{U}}\mathbf{\widetilde{Z}}}
	= 
	\frac{\mathsf{SNR}^3(\mathsf{SNR}(1-\gamma^2) + 1)}{(\mathsf{SNR}^2(1-\gamma^2) + 2\mathsf{SNR} + 1)^2}
	\begin{bmatrix}
		\mathbf{0}_{2\times 2} & -\mathbf{S}_{\gamma,\theta} \\
		-\mathbf{S}_{\gamma,\theta} & \mathbf{0}_{2\times 2}
	\end{bmatrix}
\end{equation*}
where 
\begin{equation*}
	\mathbf{S}_{\gamma,\theta}
	=
	\begin{bmatrix}
		-\gamma\cos(2\theta) & \gamma\sin(2\theta)\\
		\gamma\sin(2\theta) & \gamma\cos(2\theta)
	\end{bmatrix}\text{,}
\end{equation*}
and
\begin{multline*}
	\mathbf{K}_{\mathbf{\widetilde{Z}}\mathbf{\widetilde{Z}}}
	=\\
	\frac{\mathsf{SNR}^2(\mathsf{SNR}+1)(\mathsf{SNR}(1-\gamma^2) + 1)}{(\mathsf{SNR}^2(1-\gamma^2)+2\mathsf{SNR}+1)^2}
	\begin{bmatrix}
		\mathbf{I}_{2} & \mathbf{T}_{\gamma,\theta} \\
		\mathbf{T}_{\gamma,\theta} & \mathbf{I}_2
	\end{bmatrix}
\end{multline*}
where
\begin{equation*}
	\mathbf{T}_{\gamma,\theta}
	=
	\frac{\mathsf{SNR}^2(1-\gamma^2)-1}{(\mathsf{SNR}+1)(\mathsf{\mathsf{SNR}(1-\gamma^2)+1})}
	\mathbf{S}_{\gamma,\theta}\text{.}
\end{equation*}

We then have for $i \in \{1,2,3,4\}$
that 
\begin{align*}
	\mathsf{SNR}_i(\gamma,\theta) &= \frac{(1-\gamma^2)\mathsf{SNR}^2 + \mathsf{SNR}}{\mathsf{SNR} + 1}\\
	&\geq 
	\frac{(1-\alpha^2)\mathsf{SNR}^2 + \mathsf{SNR}}{\mathsf{SNR} + 1}\text{.}
\end{align*}
We can then achieve a rate of 
\begin{equation*}
	C\mleft(\frac{(1-\alpha^2)\mathsf{SNR}^2 + \mathsf{SNR}}{\mathsf{SNR} + 1}\mright)
	= C_\alpha(\mathsf{SNR}) - C(\mathsf{SNR})
\end{equation*}
on each sub-channel 
and can thus achieve the parallel capacity 
$$ 2C_\alpha^\mathsf{p}(\mathsf{SNR}) = 4C_\alpha(\mathsf{SNR}) 
- 4C(\mathsf{SNR})\text{.}$$

To achieve the full capacity, as before, we code across
many channel uses of the first and second sub-channels whose
SNRs satisfy 
\begin{align*}
	\mathsf{SNR}_1(\gamma,\theta) &\geq 	\frac{(1-\alpha^2)\mathsf{SNR}^2 + \mathsf{SNR}}{\mathsf{SNR} + 1}\\
	\mathsf{SNR}_2(\gamma,\theta) &\geq 	\frac{(1-\alpha^2)\mathsf{SNR}^2 + \mathsf{SNR}}{\mathsf{SNR} + 1}
\end{align*}
and then recover $U_1$ and $U_2$. We then perform
the interference cancellation and re-equalization
\textit{exactly} as in Section \ref{real-ZF-SIC}.

The SNRs seen by $U_1,U_2,U_3,U_4$ are then given by
\begin{align*}
	\mathsf{SNR}_1(\gamma,\theta) &\geq \frac{(1-\alpha^2)\mathsf{SNR}^2 + \mathsf{SNR}}{\mathsf{SNR} + 1}\\
	\mathsf{SNR}_2(\gamma,\theta) &\geq \frac{(1-\alpha^2)\mathsf{SNR}^2 + \mathsf{SNR}}{\mathsf{SNR} + 1}\\
	\mathsf{SNR}_3(\gamma,\theta) &= \mathsf{SNR}\\
	\mathsf{SNR}_4(\gamma,\theta) &= \mathsf{SNR}
\end{align*}
and we can thus achieve the full capacity \eqref{compound-capacity} by LMMSE-SIC in conjunction
with the precoder \eqref{magic-1}, i.e.,
\begin{equation*}
	2C\mleft(\frac{(1-\alpha^2)\mathsf{SNR}^2 + \mathsf{SNR}}{\mathsf{SNR} + 1}\mright)
	+
	2C(\mathsf{SNR}) = 2C_\alpha(\mathsf{SNR})\text{.}
\end{equation*}

Finally, we remark that, as before, from the 
signal--noise cross-covariance and noise covariance 
matrices, we see that there are no correlations
within the first and second sub-channels. We can
then spread a single codeword across the first
and second sub-channels and a single codeword
across the third and fourth sub-channels from
codes of different rates commensurate with
the two SNRs as before. 

\section{Optimal Scheme for Complex-Valued PDL Channels}\label{Complex-Scheme}

\subsection{Setup}
We now provide an optimal scheme for the more general case of a complex-valued
channel matrix.
In such a situation, the channel matrix is given by 
\begin{equation*}
	\begin{bmatrix}
		\sqrt{1+\gamma} & 0\\
		0 & \sqrt{1-\gamma} 
	\end{bmatrix}
	\begin{bmatrix}
		\cos(\theta) & -\sin(\theta)\\
		\sin(\theta) & \cos(\theta)
	\end{bmatrix}
	\begin{bmatrix}
		e^{i\phi} & 0 \\
		0 & e^{-i\phi}
	\end{bmatrix}
\end{equation*}
where $\gamma\in[-\alpha,\alpha]$, $\theta \in[0,2\pi)$, and $\phi\in[0,2\pi)$.
However, we will work with an equivalent
real-valued model because the proposed scheme in this case turns out to require
\textit{widely linear} processing, i.e., processing which acts \textit{differently}
on the real and imaginary parts of the signals involved. This is mathematically
equivalent to real linear processing under an equivalent real-valued description
where the first and second halves of a vector contain the real and imaginary parts
respectively. 

The equivalent real-valued model is then given by
\begin{equation*}
	\mathbf{Y} = \mathbf{D}_\gamma\mathbf{R}_\theta \mathbf{B}_\phi \mathbf{X} + \mathbf{Z}
\end{equation*}
where $\mathbf{Z}\sim\mathcal{N}(\mathbf{0},\mathbf{I}_4)$, $\mathbb{E}[\mathbf{X}\mathbf{X}^\intercal] = \mathsf{SNR} \mathbf{I}_4$, and 
\begin{align*}
	\mathbf{D}_\gamma
	&=
	\begin{bmatrix}
		\sqrt{1+\gamma} & 0 & 0 & 0 \\
		0 & \sqrt{1-\gamma} & 0 & 0 \\
		0 & 0 & \sqrt{1+\gamma} & 0 \\
		0 & 0 & 0 & \sqrt{1-\gamma}
	\end{bmatrix}\\
	\mathbf{R}_\theta
	&= 
	\begin{bmatrix}
		\cos(\theta) & -\sin(\theta) & 0 & 0 \\
		\sin(\theta) & \cos(\theta) & 0 & 0 \\
		0 & 0 & \cos(\theta) & -\sin(\theta) \\
		0 & 0 & \sin(\theta) & \cos(\theta)
	\end{bmatrix}\\
 	\mathbf{B}_\phi
 	&=
 	\begin{bmatrix}
 		\cos(\phi) & 0 & -\sin(\phi) & 0 \\
 		0 & \cos(\phi) & 0 & \sin(\phi) \\
 		\sin(\phi) & 0 & \cos(\phi) & 0 \\
 		0 & -\sin(\phi) & 0 & \cos(\phi)
 	 \end{bmatrix}\text{.}
\end{align*}
One can easily verify that the compound capacity, normalized by the number of real
dimensions, is identical to the case of a real-valued channel since the rotation is irrelevant
to the capacity calculation.

As before, consider a
two-channel-use extension of the channel
and take the input to be 
$\mathbf{X} = \mathbf{G}\mathbf{U}$ where $\mathbf{G} \in \mathbb{R}^{8\times 8}$
is an orthogonal matrix so that $\mathbb{E}[\mathbf{X}\mathbf{X}^\intercal] = \mathbb{E}[\mathbf{U}\mathbf{U}^\intercal] = \mathsf{SNR}\mathbf{I}_8$\text{.}
The effective channel matrix is then 
\begin{equation*}
	\mathbf{H} = 
	\diag\mleft(\mathbf{D}_\gamma\mathbf{R}_\theta\mathbf{B}_\phi, 
	\mathbf{D}_\gamma\mathbf{R}_\theta\mathbf{B}_\phi \mright)\mathbf{G} \in \mathbb{R}^{8 \times 8}
\end{equation*}
with the effective channel being 
$\mathbf{Y} = \mathbf{H}\mathbf{U} + \mathbf{Z}$
where $\mathbf{Z}\sim\mathcal{N}(\mathbf{0},\mathbf{I}_8)$ 
so that when $\mathbf{U}\sim \mathcal{N}(\mathbf{0},\mathsf{SNR}\mathbf{I}_8)$,
we have 
\begin{align*}
	4C_\alpha(\mathsf{SNR})
	&= 
	\min_{\gamma,\theta,\phi}
	I(\mathbf{U};\mathbf{Y})\\
	&=\min_{\gamma,\theta,\phi}
	\left[\sum_{i=1}^{8} I(U_i;\mathbf{Y}U_1\cdots U_{i-1})\right]\text{.}
\end{align*}

As before, it suffices to find a precoder $\mathbf{G}$ such that
\begin{multline*}
\min_{\gamma,\theta,\phi}
	\left[\sum_{i=1}^{8} I(U_i;\mathbf{Y}U_1\cdots U_{i-1})\right]\\
	=
	\sum_{i=1}^{8} \min_{\gamma,\theta,\phi} I(U_i;\mathbf{Y}U_1\cdots U_{i-1})\text{.}
	\tag{$\star\star$}
\end{multline*}

Surprisingly, such a precoder exists again
and is given by
\begin{equation}
	\mathbf{G}
	=
	\frac{1}{\sqrt{2}}
	\begin{bmatrix}
		1 & 0 & 0 & 0 & 1 & 0 & 0 & 0\\ 0 & 0 & 1 & 0 & 0 & 0 & 1 & 0\\ 0 & -1 & 0 & 0 & 0 & -1 & 0 & 0\\ 0 & 0 & 0 & -1 & 0 & 0 & 0 & -1\\ 0 & 0 & 1 & 0 & 0 & 0 & -1 & 0\\ -1 & 0 & 0 & 0 & 1 & 0 & 0 & 0\\ 0 & 0 & 0 & 1 & 0 & 0 & 0 & -1\\ 0 & -1 & 0 & 0 & 0 & 1 & 0 & 0 
	\end{bmatrix}\label{magic-2}\text{.}
\end{equation}
We will demonstrate that \eqref{magic-2} satisfies the stronger property
\begin{multline*}
	\sum_{i=1}^{4} \min_{\gamma,\theta,\phi} I(U_i;\mathbf{Y})
	+
	\sum_{i=5}^{8} \min_{\gamma,\theta,\phi} I(U_i;\mathbf{Y}U_1U_2U_3U_4)\\
	= 4C_\alpha(\mathsf{SNR})
\end{multline*}
and thus satisfies ($\star\star$).

As before, we will begin with ZF-SIC for simplicity but
we will be less detailed since the scheme is similar to
that in Section \ref{Real-Scheme}.

\subsection{ZF-SIC}

By taking $\mathbf{E} = \mathbf{H}^{-1}$ and 
proceeding as in Section \ref{SNR-Calculation-Section},
one finds that
\begin{align*}
	\mathbf{K}_{\widetilde{\mathbf{U}}\widetilde{\mathbf{U}}} 
	&= \mathsf{SNR}\mathbf{I}_8\\
	\mathbf{K}_{\widetilde{\mathbf{U}}\widetilde{\mathbf{Z}}} 
	&= \mathbf{0}_{8\times 8}
\end{align*}
as expected and that
\begin{equation*}
	\mathbf{K}_{\widetilde{\mathbf{Z}}\widetilde{\mathbf{Z}}} 
	=
	\frac{1}{1-\gamma^2}
	\begin{bmatrix}
		\mathbf{I}_4 & \mathbf{S}_{\gamma,\theta,\phi} \\
		\mathbf{S}_{\gamma,\theta,\phi} & \mathbf{I}_4
	\end{bmatrix}
\end{equation*} 
where $\mathbf{S}_{\gamma,\theta,\phi} \in \mathbb{R}^{4\times 4}$ is a matrix
depending on $\gamma$, $\theta$, and $\phi$ which we will not bother to write out explicitly. 
We then have for $i \in \{1,2,3,4,5,6,7,8\}$
that 
\begin{align*}
	\mathsf{SNR}_i(\gamma,\theta,\phi) &= (1-\gamma^2)\mathsf{SNR} \\ 
	&\geq (1-\alpha^2)\mathsf{SNR}\text{.}
\end{align*}

We then assume that $U_1,U_2,U_3,U_4$ have been recovered 
and proceed to the interference cancellation.
One can verify that we have
\begin{equation*}
	\mathbf{H}^\intercal\mathbf{H} = 
	\begin{bmatrix}
		\mathbf{I}_4 & -\mathbf{S}_{\gamma,\theta,\phi}\\
		-\mathbf{S}_{\gamma,\theta,\phi} & \mathbf{I}_4
	\end{bmatrix}\text{.}
\end{equation*}
so that we can partition $\mathbf{H}$ as
\begin{equation*}
	\mathbf{H} = 
	\begin{bmatrix}
		\mathbf{H}_1 & \mathbf{H}_2
	\end{bmatrix}
\end{equation*}
where $\mathbf{H}_1,\mathbf{H}_2 \in \mathbb{R}^{8\times 4}$ and satisfy
$\mathbf{H}_1^\intercal\mathbf{H}_1 = \mathbf{H}_2^\intercal\mathbf{H}_2 = \mathbf{I}_4$.
Cancelling the interference from $U_1,U_2,U_3,U_4$ then yields
the effective channel
\begin{equation*}
	\mathbf{Y}-\mathbf{H}_1\begin{bmatrix}
		U_1 \\
		U_2 \\
		U_3 \\
		U_4
	\end{bmatrix} 
	= \mathbf{H}_2\begin{bmatrix}
		U_5 \\
		U_6 \\
		U_7 \\
		U_8
	\end{bmatrix} 
	+
	\mathbf{Z}
\end{equation*}
which we can optimally equalize with $\mathbf{E}=\mathbf{H}_2^\intercal$ to get
\begin{equation*}
	\mathbf{\widehat{Y}} = 
	\mathbf{H}_2^\intercal\left(\mathbf{Y}-\mathbf{H}_1\begin{bmatrix}
		U_1 \\
		U_2 \\
		U_3 \\
		U_4 
	\end{bmatrix}\right) 
	= \underbrace{\begin{bmatrix}
			U_5 \\
			U_6 \\
			U_7 \\
			U_8
	\end{bmatrix}}_{\mathbf{\widehat{U}}} + \underbrace{\mathbf{H}_2^\intercal\mathbf{Z}}_{\mathbf{\widehat{Z}}}
\end{equation*}
which yields
\begin{align*}
	\mathbf{K}_{\widehat{\mathbf{U}}\widehat{\mathbf{U}}} 
	&= \mathsf{SNR}\mathbf{I}_4\\
	\mathbf{K}_{\widehat{\mathbf{U}}\widehat{\mathbf{Z}}} 
	&= \mathbf{0}_{4\times 4}\\
	\mathbf{K}_{\widehat{\mathbf{Z}}\widehat{\mathbf{Z}}} 
	&= \mathbf{I}_{4}\text{.}
\end{align*}

The SNRs seen by $U_1,U_2,U_3,U_4,U_5,U_6,U_7,U_8$ are then given by
\begin{align*}
	\mathsf{SNR}_1(\gamma,\theta,\phi) &\geq (1-\alpha^2)\mathsf{SNR}\\
	\mathsf{SNR}_2(\gamma,\theta,\phi) &\geq (1-\alpha^2)\mathsf{SNR}\\
	\mathsf{SNR}_3(\gamma,\theta,\phi) &\geq (1-\alpha^2)\mathsf{SNR}\\
	\mathsf{SNR}_4(\gamma,\theta,\phi) &\geq (1-\alpha^2)\mathsf{SNR}\\
	\mathsf{SNR}_5(\gamma,\theta,\phi) &= \mathsf{SNR}\\
	\mathsf{SNR}_6(\gamma,\theta,\phi) &= \mathsf{SNR}\\
	\mathsf{SNR}_7(\gamma,\theta,\phi) &= \mathsf{SNR}\\
	\mathsf{SNR}_8(\gamma,\theta,\phi) &= \mathsf{SNR}
\end{align*}
and we can thus achieve under ZF-SIC and precoding with \eqref{magic-2},
$$4\widetilde C_\alpha(\mathsf{SNR}) = 4C((1-\alpha^2)\mathsf{SNR}) + 4C(\mathsf{SNR})$$
which is within $\mathcal{O}(\mathsf{SNR}^{-1})$ of the true capacity $4C_\alpha(\mathsf{SNR})$. 

\subsection{LMMSE-SIC}

As before, achieving the full
capacity is merely a matter of replacing
the ZF equalizer with the LMMSE equalizer
in the ZF-SIC procedure just described. By taking 
\begin{equation*}
	\mathbf{E} =
	\mathbf{H}^\intercal\left(\mathbf{H}\mathbf{H}^\intercal + \frac{1}{\mathsf{SNR}}\mathbf{I}_4\right)^{-1}
\end{equation*}
and proceeding as in Section \ref{SNR-Calculation-Section}, we get
\begin{equation*}
	\mathbf{K}_{\mathbf{\widetilde{U}}\mathbf{\widetilde{U}}}
	=
	\frac{\mathsf{SNR}^3(\mathsf{SNR}(1-\gamma^2) + 1)^2}{(\mathsf{SNR}^2(1-\gamma^2) + 2\mathsf{SNR} + 1)^2}
	\mathbf{I}_8\text{.}
\end{equation*}
Moreover, we get
\begin{equation*}
	\mathbf{K}_{\mathbf{\widetilde{U}}\mathbf{\widetilde{Z}}}
	= 
	\frac{\mathsf{SNR}^3(\mathsf{SNR}(1-\gamma^2) + 1)}{(\mathsf{SNR}^2(1-\gamma^2) + 2\mathsf{SNR} + 1)^2}
	\begin{bmatrix}
		\mathbf{0}_{4\times 4} & -\mathbf{S}_{\gamma,\theta,\phi} \\
		-\mathbf{S}_{\gamma,\theta,\phi} & \mathbf{0}_{4\times 4}
	\end{bmatrix}
\end{equation*}
and 
\begin{multline*}
	\mathbf{K}_{\mathbf{\widetilde{Z}}\mathbf{\widetilde{Z}}}
	=\\
	\frac{\mathsf{SNR}^2(\mathsf{SNR}+1)(\mathsf{SNR}(1-\gamma^2) + 1)}{(\mathsf{SNR}^2(1-\gamma^2)+2\mathsf{SNR}+1)^2}
	\begin{bmatrix}
		\mathbf{I}_{4} & \mathbf{T}_{\gamma,\theta,\phi} \\
		\mathbf{T}_{\gamma,\theta,\phi} & \mathbf{I}_4
	\end{bmatrix}
\end{multline*}
where
\begin{equation*}
	\mathbf{T}_{\gamma,\theta,\phi}
	=
	\frac{\mathsf{SNR}^2(1-\gamma^2)-1}{(\mathsf{SNR}+1)(\mathsf{\mathsf{SNR}(1-\gamma^2)+1})}
	\mathbf{S}_{\gamma,\theta,\phi}\text{.}
\end{equation*}

We then have for $i \in \{1,2,3,4,5,6,7,8\}$
that 
\begin{align*}
	\mathsf{SNR}_i(\gamma,\theta) &= \frac{(1-\gamma^2)\mathsf{SNR}^2 + \mathsf{SNR}}{\mathsf{SNR} + 1}\\
	&\geq 
	\frac{(1-\alpha^2)\mathsf{SNR}^2 + \mathsf{SNR}}{\mathsf{SNR} + 1}\text{.}
\end{align*}
Exactly as before, $U_1,U_2,U_3,U_4$ will see this SNR instead of $(1-\alpha^2)\mathsf{SNR}$
and the interference cancellation and re-equalization step remains the same
so that $U_5,U_6,U_7,U_8$ see an SNR of $\mathsf{SNR}$. We then achieve, using LMMSE-SIC and the precoder
\eqref{magic-2}, the
full capacity 
\begin{equation*}
	4C\mleft(\frac{(1-\alpha^2)\mathsf{SNR}^2 + \mathsf{SNR}}{\mathsf{SNR} + 1}\mright)
	+
	4C(\mathsf{SNR}) = 4C_\alpha(\mathsf{SNR})\text{.}
\end{equation*}

Finally, we note that there are no correlations within
the first, second, third, and fourth sub-channels, 
and no correlations within the fifth, sixth, seventh, and eighth 
sub-channels. Therefore, as before, we can combine
these groups and only have to send two codewords from 
two codes of different rates.

\section{Performance and Practical Considerations}\label{Practical}

\subsection{Performance}

While we will defer a fine-grained and practically-minded analysis 
of the proposed scheme to future work, a coarse-grained analysis
is immediately possible.
In particular, since the scheme is entirely a reduction to classical AWGN communication,
no new simulations are necessary to determine the performance from an FER and gap-to-capacity perspective. 
We will consider the case of ZF-SIC for simplicity, but identical reasoning applies to LMMSE-SIC. 
We require two coded modulation schemes with one
operating at a rate of 
\begin{equation*}
	C\mleft(\frac{(1-\alpha^2)\mathsf{SNR}}{\mathsf{g}_1}\mright)
\end{equation*}
bits per real dimension 
and the other operating at a rate of
\begin{equation*}
	C\mleft(\frac{\mathsf{SNR}}{\mathsf{g}_2}\mright)
\end{equation*}
bits per real dimension where $\mathsf{g}_1$ and $\mathsf{g}_2$
are the respective gaps to capacity.
At high SNR, the overall gap to the compound channel capacity $C_\alpha(\mathsf{SNR})$ is then given by $\sqrt{\mathsf{g}_1\mathsf{g}_2}$
or
\begin{equation}\label{gap-to-capacity}
	\frac{1}{2}\big[10\log_{10}(\mathsf{g}_1) + 10\log_{10}(\mathsf{g}_2)\big] \text{ dB}\text{.}
\end{equation}
Relative to the classical AWGN capacity $2C(\mathsf{SNR})$, we have an additional
gap of \eqref{sic-penalty} which is the fundamental cost of PDL and is as small
as theoretically possible.

Next, we consider the performance from an FER perspective. Suppose that
we have FER data for the real classical AWGN performance of the two constituent 
coded modulation schemes given by $\mathsf{FER}_1(\mathsf{SNR})$
and $\mathsf{FER}_2(\mathsf{SNR})$. The overall FER denoted $\mathsf{FER}(\mathsf{SNR})$
is determined as follows. Denote by $D_1$ the event that the first
codeword is decoded correctly and denote by $D_2$ the event that the second
codeword is decoded correctly. Under the proposed scheme, the first codeword
sees an SNR of $(1-\alpha^2)\mathsf{SNR}$ and the second codeword
sees an SNR of $\mathsf{SNR}$ provided that the first codeword is
decoded correctly. We then have
\begin{align*}
	\Pr(D_1) &= 1 - \mathsf{FER}_1((1-\alpha^2)\mathsf{SNR})\\
		\Pr(D_2 \mid D_1) &= 1 - \mathsf{FER}_2(\mathsf{SNR})\text{.}
\end{align*}
Denoting by $D$ the event that the overall frame is decoded 
correctly, we have 
\begin{equation*}
	\Pr(D) = \Pr(D_1 \cap D_2) = \Pr(D_1) \Pr(D_2 \mid D_1) 
\end{equation*} 
and $\mathsf{FER}(\mathsf{SNR}) = 1-\Pr(D)$.
This yields
\begin{align}
		&\mathsf{FER}(\mathsf{SNR})\nonumber\\ &=  1 - (1 - \mathsf{FER}_1((1-\alpha^2)\mathsf{SNR}))(1 - \mathsf{FER}_2(\mathsf{SNR}))\nonumber \\
		&= \mathsf{FER}_1((1-\alpha^2)\mathsf{SNR}) + \mathsf{FER}_2(\mathsf{SNR})\nonumber \\ 
		&\phantom{=}- \mathsf{FER}_1((1-\alpha^2)\mathsf{SNR})\cdot\mathsf{FER}_2(\mathsf{SNR}) \nonumber\\
		&\leq
		\mathsf{FER}_1((1-\alpha^2)\mathsf{SNR}) + \mathsf{FER}_2(\mathsf{SNR})\text{,} \label{FER-bound}
\end{align}
with the upper bound \eqref{FER-bound} being a good estimate since the product term is typically 
negligible.

One can then substitute FER and gap-to-capacity data from off-the-shelf 
schemes such as those in \cite{bocherer-cookbook,masoud-cookbook} into 
\eqref{gap-to-capacity} and \eqref{FER-bound} to determine the performance
of the proposed scheme. 
However, one will not necessarily be able to find perfectly
rate-matched code pairs given the unconventional requirement of
two codes with a certain rate gap under this scheme. We suggest the construction
of such code pairs for different SNRs (or overall rates) as a problem for future work. 

\subsection{A Concrete Example}\label{Concrete}

We now provide a concrete instantiation of the proposed scheme by considering
a particular practical coded modulation scheme from \cite{bocherer-cookbook}. 
The schemes provided in \cite{bocherer-cookbook} 
entail bit-interleaved coded modulation (BICM) with standard low-density parity-check (LDPC) codes 
and probabilistic amplitude shaping (PAS) with standard bipolar amplitude shift keying (ASK) constellations. The schemes are designed to operate within around $1$ dB of the classical
AWGN capacity and we necessarily expect to achieve comparable gaps to the compound PDL channel capacity when combining them with the proposed scheme as per \eqref{gap-to-capacity}.

Suppose that we have a worst-case PDL of $6$ dB, i.e., $\alpha = 0.599$,
and an SNR of $10\log_{10}(\mathsf{SNR}) = 13.01 \text{ dB}$, i.e., $\mathsf{SNR} = 20$.
The two coded modulation schemes we consider in this case are 
\begin{itemize}
	\item $8$-ASK, a rate $3/4$ LDPC code, and PAS leading to a rate of $1.8$ bits per real symbol (see \cite[Table~IX]{bocherer-cookbook}); and
	\item $16$-ASK, a rate $5/6$ LDPC code, and PAS leading to a rate of $2.1$ bits per real symbol (see \cite[Table~VIII]{bocherer-cookbook}). 
\end{itemize}
We then form vectors of coded symbols where the first and second halves of the entries
of the vectors are coded symbols from these two coded modulation schemes respectively, 
precode as per \eqref{magic-1} or \eqref{magic-2}, and perform the proposed ZF-SIC 
procedure with bit-metric LDPC decoding as in \cite{bocherer-cookbook}.

Noting that $10\log_{10}((1-0.599^2)\cdot 20) = 11.08 \text{ dB}$ and referring to the data in \cite[Table~IX]{bocherer-cookbook}) and  \cite[Table~VIII]{bocherer-cookbook} respectively, we have that 
\begin{align*}
	\mathsf{FER}_1((1-0.599^2)\cdot 20) &= 1.2\cdot 10^{-3}\\
	\mathsf{FER}_2(20) &= 1.3\cdot 10^{-3}
\end{align*}
which yields 
\begin{equation*}
	\mathsf{FER}(20) \leq 2.5\cdot 10^{-3}
\end{equation*}
where this is also an upper bound on the bit error rate (BER).
Moreover, the overall transmission rate is $(1.8+2.1)/2 = 1.95$ bits per real dimension.
This operating point is plotted in Figure \ref{Capacities-Figure-Zoomed} as a solid circle where we see a gap to capacity of under $0.7$ dB. Note that the translation from bits per real dimension to nominal spectral efficiency in bits per second per
hertz is multiplication by a factor of four since we have two real symbols (or one complex symbol) per second per hertz per polarization and two polarizations. Moreover, we expect some further rate loss due to additional outer
FEC which would be needed to bring the BER down to, e.g., $10^{-15}$.

\subsection{Complexity}

The proposed scheme is, in some sense, as simple as possible since we 
are given a channel with two polarizations having two different capacities
and we synthesize by linear processing and one interference cancellation 
step, two scalar channels with two different capacities. However, these
new capacities are rotation-independent allowing us to code separately
across them provided that we decode successively.

In \cite{dumenil-fundamental}, an empirically near-capacity scheme is reported
under identical modelling assumptions and is 
based on concatenation of a spatially-coupled LDPC code
with a Silver code and joint processing of the two polarizations via iterative demapping and equalization. 
In contrast, the proposed scheme has:
\begin{itemize}
	\item a \textit{linear} equalizer using only \textit{one} iteration of decision feedback, 
	\item significantly smaller peak-to-average power ratio (PAPR) since the precoding is essentially the same as \cite{oyama,pairwise},
	\item and provable optimality with the performance only
	limited by the classical AWGN performance of the constituent codes.
\end{itemize}
The primary disadvantages of the proposed scheme are the need for large memory to store the first codeword after decoding, as well as the need for two codes of different rates and two corresponding decoders. In principle, this second disadvantage should not be a fundamental issue since the overall
throughput is split between the decoders. Moreover, the only way to bypass this issue without loss of optimality
would be with joint coding and joint decoding which is inherently more complex than separate coding with separate but successive decoding.

\subsection{Practical Considerations}

We note that under various architectural 
assumptions, the proposed scheme remains optimal or applicable even if 
our mathematical assumptions do not apply to the physical channel. 
For example,
given the output of a standard blind adaptive equalizer used for joint 
polarization mode dispersion (PMD) and state of polarization (SOP)
compensation, one recovers a pair of parallel AWGN channels
with correlated noise and SNR imbalance. 
Upon covariance estimation and noise whitening, one recovers
precisely the communication scenario considered in this paper.
As a further example, it was recently  
demonstrated in \cite{USIC_1, USIC_2, USIC_3} 
that coded LMMSE-SIC schemes, similar to the proposed one,
can be made to work in practice with standard blind adaptive equalizers, eliminating the need 
for explicit channel estimation.

\section{Conclusion}\label{Conclusion}

We have demonstrated that information-theoretically optimal
PDL mitigation is obtained by simply using LMMSE-SIC 
in conjunction with an appropriate special choice of precoder. While experimental 
validation has been deferred to future work, the underlying architecture
is standard and highly amenable to practical implementation. Moreover, we
expect a $1$ dB gain over PDL mitigation schemes based on parallel
architectures at the very small cost of a single post-FEC interference cancellation. 
On the other hand, we expect significantly lower complexity than other schemes 
having comparable performance.

\appendix[On the Generalizability of the Proposed Scheme]
	The precoding matrices \eqref{magic-1} and \eqref{magic-2} were pulled out of a hat so it is natural
	to ask whether there is an underlying principle to their construction and whether it generalizes to higher-dimensional SDM systems with MDL. Unfortunately, while there is indeed an underlying principle, we expect that generalizations to higher dimensions are not possible or are suboptimal since the scheme hinges on a multitude of coincidences. 
	These are that 
	\begin{itemize}
		\item the arithmetic average of the squares of the channel singular values is $1$;
		\item $2\times 2$ and $4 \times 4$ orthogonal designs exist \cite{orthogonal-designs-book,orthogonal-designs-paper}; and
		\item the arithmetic, geometric, and harmonic averages, $\mathsf{A}$, $\mathsf{G}$, and $\mathsf{H}$, respectively, of \textit{two} positive numbers satisfy
		\begin{equation}\label{GGAH}
			\mathsf{G}^2 = \mathsf{A}\cdot \mathsf{H}\text{.}
		\end{equation}
	\end{itemize}
	We will proceed to elaborate on this. 
	
	Suppose that we have a channel matrix with positive singular values $\sqrt{a}$ and $\sqrt{b}$.
	At high SNR and under a symmetric power allocation, the capacity is essentially
	\begin{align}
		&\frac{1}{2}\log_2(a\mathsf{SNR}) + \frac{1}{2}\log_2(b\mathsf{SNR})\nonumber\\
		&= \frac{1}{2}\log_2(ab\mathsf{SNR}^2)\label{cap-is-product}\\
		&= \frac{1}{2}\log_2(\sqrt{ab}\mathsf{SNR}) + \frac{1}{2}\log_2(\sqrt{ab}\mathsf{SNR})
		\label{GM2}\\
		&= \frac{1}{2}\log_2\mleft(\frac{a+b}{2}\mathsf{SNR}\mright) 
		+ \frac{1}{2}\log_2\mleft(\frac{2ab}{a+b}\mathsf{SNR}\mright) \label{AH}
	\end{align} 
	where the equality between \eqref{GM2} and \eqref{AH} is essentially a statement of 
	the identity \eqref{GGAH}. 
	
	Observe from \eqref{cap-is-product} that the compound (or worst-case)
	capacity is determined by the minimum value of the product $ab$ over the set of values that $a$ and $b$ are allowed to take. This means that if we can
	precode so that ZF-SIC synthesizes sub-channels with gains that are equal to the geometric 
	mean of the squares of the singular values, we can achieve the compound capacity with ZF-SIC since the minimum 
	of \eqref{cap-is-product} would be equal to the sum of the minima of the terms in \eqref{GM2}, minimizing over the possible values for $a$ and $b$. 
	
	Such a precoder can be found by using a geometric mean decomposition (GMD) \cite{GMD} of the channel matrix but will depend on the channel matrix and thus require channel knowledge at the transmitter. 
	Suppose, on the other hand, that we find a precoder which synthesizes sub-channels having gains equal to the arithmetic and harmonic averages of the squares of the channel singular values respectively. This is not guaranteed to result in optimality of ZF-SIC because the sum of the minima of the terms
	in \eqref{AH} need not necessarily coincide with the minimum of \eqref{cap-is-product} over all admitted values of $a$ and $b$ \textit{unless} all of those values satisfy $a+b = c$ for some constant $c$, i.e., lie on a line. This is indeed the case for the channel model considered in this paper in which $a = 1+\gamma$ and $b = 1-\gamma$ so that $(a+b)/2 = 1$.
	
	It then remains to construct a \textit{channel-independent} precoder which produces sub-channels with harmonic and arithmetic average means under ZF-SIC. While it is generally possible to construct
	channel-independent precoders which produce harmonic average gains under ZF as done in \cite{damen} for SDM channels with MDL for an arbitrary number of dimensions, we further require arithmetic average gains after interference cancellation. This would be guaranteed by \eqref{AH} and the chain rule of mutual information if the precoding was across a single channel use, but any channel-independent precoder must act across multiple channel uses else the channel could simply undo the precoding. Thus we seek 
	a channel-independent precoder, necessarily across multiple channel uses, such that we see harmonic average gains under ZF and arithmetic 
	average gains after interference cancellation. This would occur if the effective channel matrix were unconditionally orthogonal after interference cancellation, i.e., had unconditionally orthogonal fixed submatrices.
	
	We can accomplish this by exploiting the existence of certain \textit{orthogonal designs}.
	An \textit{orthogonal design} is a matrix in indeterminates which is unconditionally orthogonal for any choice of these indeterminates.  Famously, only $2\times 2$, $4 \times 4$, and $8\times 8$ orthogonal designs exist \cite{orthogonal-designs-paper,orthogonal-designs-book}.
	In the case of a real-valued channel, we choose a precoder which induces a $2\times 2$ orthogonal design structure in the effective channel matrix with the columns of the original channel matrix playing the role of
	the indeterminates resulting in unconditionally orthogonal $4 \times 2$ submatrices. In the case of a complex-valued channel, we similarly induce a $4\times 4$ orthogonal design structure resulting in unconditionally orthogonal $8 \times 4$ submatrices.  
	
	While orthogonal designs have been famously used to construct an astonishingly vast variety of space--time coding schemes, 
	the particular scheme and analysis occurring in this paper does not occur in the wireless communications literature---to the best of the authors' knowledge---because the $(a+b)/2 = 1$ assumption does not occur in the modelling of wireless fading channels. This assumption represents
	the absence of IL uncertainty or the presence of perfect dynamic gain equalization which cannot be realized in a wireless setting where $(a+b)/2$ is highly random as opposed to deterministic or tightly concentrated around its mean. 
	
	Finally, one might wonder if the $8\times 8$ orthogonal design can provide a generalization of the proposed scheme to $8$-mode SDM systems with MDL, but even the relationship between means \eqref{GGAH} fails to hold for more than two numbers so we do not expect the resulting scheme to be optimal. 
	The proposed scheme thus maximally exploits the structure of the problem along with sporadic mathematical constructions to achieve a simplicity which is likely not possible for generalizations of the problem.


\end{document}